# OPTIMIZATION OF STOCHASTIC DATABASE CRACKING

Meenesh Bhardwaj
Amity university
Meenesh@student.amity.edu

Mrs. Aarti Chugh
Department of Computer Science
achugh@ggn.amity.edu

## Abstract:

Variant Stochastic cracking is a significantly more resilient approach to adaptive indexing. It showed [1]that Stochastic cracking uses each query as a hint on how to reorganize data, but not blindly so; it gains resilience and avoids performance bottlenecks by deliberately applying certain arbitrary choices in its decision making. Therefore bring, adaptive indexing forward to a mature formulation that confers the workload-robustness that previous approaches lacked. Original cracking relies on the randomness of the workloads to converge well. [2][3] However, where the workload is non-random, cracking needs to introduce randomness on its own. Stochastic Cracking clearly improves over original cracking by being robust in workload changes while maintaining all original cracking features when it comes to adaptation.

But looking at both types of cracking, it conveyed an incomplete picture as at some point of time it is must to know whether the workload is random or sequential.

In this paper our focus is on optimization of variant stochastic cracking, that could be achieved in two ways either by reducing the initialization cost to make stochastic cracking even more transparent to the user, especially for queries that initiate a workload change and hence incur a higher cost or by combining the strengths of the various stochastic cracking algorithms via a dynamic component that decides which algorithm to choose for a query on the fly. The efforts have been put in to make an algorithm that reduces the initialization cost by using the main notion of both cracking, while considering the requirements of adaptive indexing [2].

## Keywords:
{Variant stochastic cracking, Optimization}

## General Terms:
{Physical organization, Workload, Cracking}

# Introduction:

Database cracking is a new query processing paradigm and adaptation paradigm towards truly self-tuned systems. Cracking requires zero human input, no a priori workload knowledge and no idle time to prepare. The ultimate goal of database cracking is to build the first truly self-organizing database system that will continuously and automatically adapt to workload changes (random). Cracking completely removes the need for human administration. Though cracking is not an auto-tuning tool, i.e., it is not an external piece of software/hardware to help with system administration. Instead cracking represents a new internal kernel design by introducing new ways of storing and accessing data. This way, the very way data is stored and subsequently accessed by queries is continuously changing to adapt to the workload and to converge to the ultimate performance.[2]

Database cracking follows both automatic index selection and partial indexes for future queries, it refined until sequential searching a partition is faster than binary searching into the AVL tree guiding a search to apply partition.[4]

## NOTION BEHIND BIG PICTURE

DB cracking is being used in Monet DB system [6][8] at the "Centrum Wiskunde & Informatica (CWI) in Amsterdam database architectures research group since 1993. Monet DB is an open-source column-store DBMS with multiple innovations in its core design. Till now many releases of Monet DB has been introduced in market i.e. Monet DB 2009, Monet DB/X Query [3][5][7].

It becomes a revolutionary DB as it stores each attribute column wise instead of row wise in traditional databases and then cracks it for the benefits of future query in terms of fast response time and throughput.

Database cracking uses THREE PIECE CRACK and TWO PIECE CRACK algorithm on "copy of a column" for very first query and rest Queries respectively; maintaining the benefits of (a) Original column remains intact (b) No overhead of maintaining complete table.

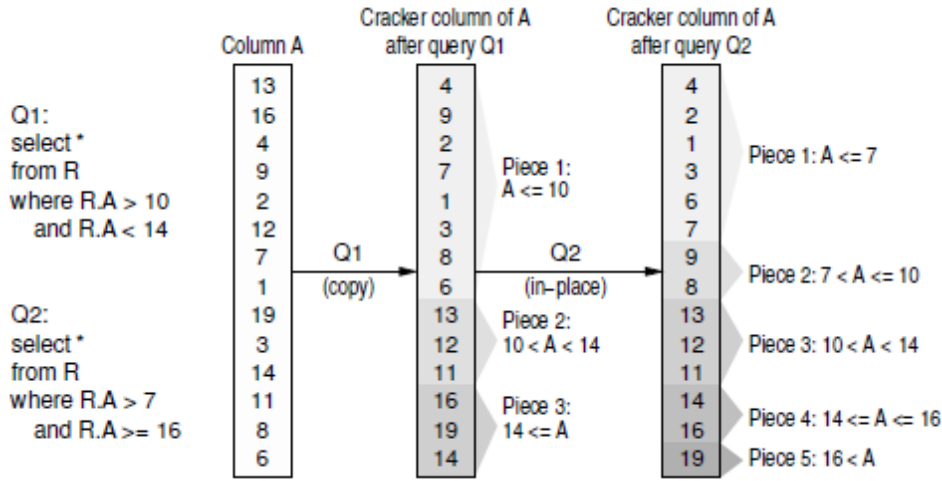

**Figure 1 showing basic Cracking method for range queries [1] [3].**

As figure shows that on copied column three piece crack algorithm is used resulting in three pieces ranging A<=10,10<A<14,14<=A and next range predicate Two piece crack algorithm has been applied resulting in 5 more pieces.

This first ever made purely column oriented database gives an extra ordinary performance when compared to traditional FULL SORTING and SCAN in dynamic environment along with benefits such as

- Performance in Random workload [3]
- Self-organizing Tuple Reconstruction in Column-stores [6]
- Self tuning without DBA[3]
- Histogram for free[7]
- No need for idle time and prior knowledge regarding workload [3][9-10]
- Deals with only required column/tables/range/queries [5][9]

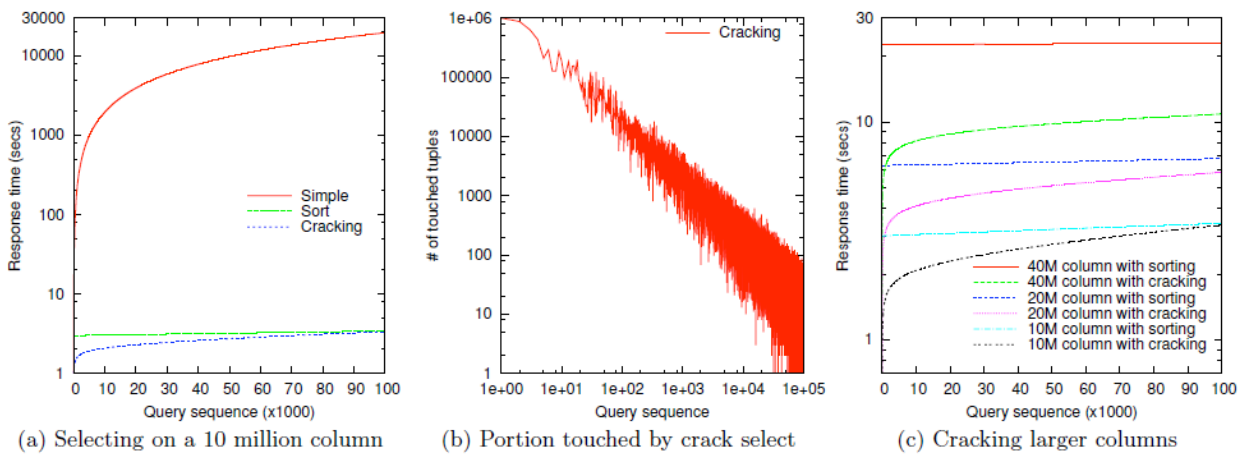

**Figure 2: shows DB cracking performance against SCAN and SORT for response time per query. [3].**

# Problem Statement:

Database Cracking is technique of Adaptive indexing that is apt for random workload inherited in dynamic environment where there is No idle and might be not enough resources to use simultaneously and continuously changing workload. It provides till now efficient performance for dynamic or random workload but it's performance degraded when workload changes from random to non random workload .[1]
As a result another approach called stochastic database cracking was introduced as a variant of database cracking that emphasized on performance improvement in non random workload.
But both techniques were for specific workload types resulting in an overall incomplete picture for real diverse workloads, as shown in figure 3.

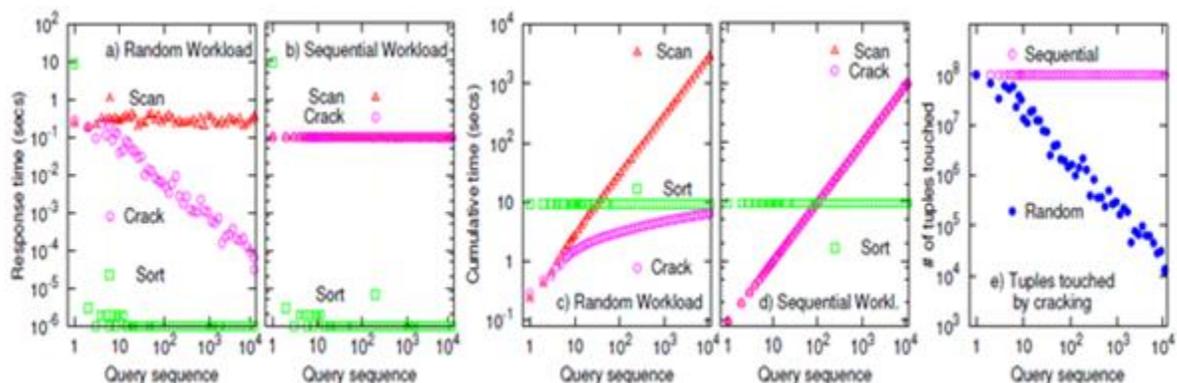

Figure 3 Basic cracking performance. Per query costs (a,b). Cumulative costs (c,d). Tuples touched (e).[1]

# Literature Review [1]

In original database cracking, cracking treats each query as a hint on how to reorganize data in a blinkered manner; it takes each query as a literal instruction on what data to index, without looking at the bigger picture. It is thanks to this literalness that cracking can instantly adapt to a random workload; yet, this literal character can also be a liability. With a non-ideal workload, strictly adhering to the queries and reorganizing the array so as to collect the query result, and only that, in a contiguous area, amounts to an inefficient quick sort-like operation; small successive portions of the array are clustered, one after the other, while leaving the rest of the array unaffected.

To solve this, Stochastic cracking ventured to drop the strict requirement in original cracking that each individual query be literally interpreted as a re-organization suggestion. It forced reorganization actions that are not strictly driven by what a query requests, but are still beneficial for the workload at large. Therefore partially driven action "by what queries want" "partially arbitrary in character".

Stochastic database cracking is a significantly more resilient approach to adaptive indexing. Stochastic cracking also uses each query as a hint on how to reorganize data, but not blindly so; it gains resilience and avoids performance bottlenecks by deliberately applying certain arbitrary choices in its decision making. Thereby, we bring adaptive indexing forward to a mature formulation that confers the workload-robustness previous approaches lacked. It has verified that stochastic cracking maintains the desired properties of original database cracking while at the same time it performs well with diverse realistic workloads. [1], while maintaining original properties of database cracking.

Stochastic cracking adopted four different techniques that try to strike a balance between
(a) Adding auxiliary reorganization steps with each query, and
(b) Remaining lightweight enough so as to significantly (if at all) not penalize individual queries.

**The algorithms used for stochastic cracking are as following**
**1) Data driven Center (DDC)**
**2) Data driven Random (DDR)**
**3) Variant of DDC and DDR (DD1C and DD1R)**
**4) Materialization data driven random1 (MDD1R)**
**5) Progressive Stochastic Cracking (PMDD1R)**

Algorithms proposed for stochastic cracking replaced the original cracking algorithms [3][10]
The overview of algorithms is given so, as to have a notion about them.

## Data Driven Center Algorithm

Data Driven Center algorithm (DDC), exercises its own decision-making without using random elements; ideally, each reorganization action should split the respective array piece in half, in a quick sort-like fashion. DDC recursively halves relevant pieces on its way to the requested range, introducing several new pieces with each new query, especially for the first queries that touch a given column. The term "Center" in its name denotes that it always tries to cut pieces in half.
DDC takes the data into account. Regardless of what kind of query arrives, DDC always performs specific data-driven actions, in addition to query driven actions. The query-driven mentality is maintained, as otherwise the algorithm would not provide good adaptation.
The DDC divide the query initially in to half no matter where its range predicate lies then only collects all qualifying tuples in a piece of [low, high], as original cracking does. But it has some limitations such as (a) adds extra operations;(b) finding these medians is an expensive and data-dependent operation;(c) it burdens individual queries with high and unpredictable costs.

## Data Driven Random Algorithm

DDR can be thought of as a single-branch quick sort. Like quick sort, it splits a piece in two, but, unlike quick sort, it only recourses into one of the two resulting pieces. The choice of that piece is again query-driven, determined by where the requested values fall.

The Algorithm is same for DDC except instead of choosing correct median for split it chooses random pivot element. It also suffers from weaknesses such as (a) adds extra operations ;(b) in worst case scenario, DDR degenerate to O(N2) cost.

## Algorithms for DD1C and DD1R (Variant of DDC and DDR)

By recursively applying more and more reorganization, both DDC and DDR managed to introduce indexing information that is useful for subsequent queries. Nevertheless, this recursive reorganization may cause the first few queries in a workload to suffer a considerably high overhead in order to perform these auxiliary operations. But an adaptive indexing solution should keep the cost of initial queries low [4][12].Therefore variants of DDC and DDR was introduced DD1C and DD1R. These variants perform at most one auxiliary physical reorganization operation. Algorithm DD1C, which works as DDC, with the difference that, after cutting a piece in half, it simply cracks the remaining piece where the requested value is located regardless of its size.

Likewise, algorithm DD1R works as DDR, but performs only one random reorganization operation before it resorts to plain cracking. Pseudo code for algorithm is same except "while "statement in Line 7 is replaced by an" if statement".

This new improved algorithm also leads to weaknesses such as (a) Reduce the initialization overhead of their recursive siblings by performing only one auxiliary reorganization operation, instead of multiple recursive ones;(b)This at most one auxiliary operation incurred high initialization cost penalized for each individual query especially when new query arrives.

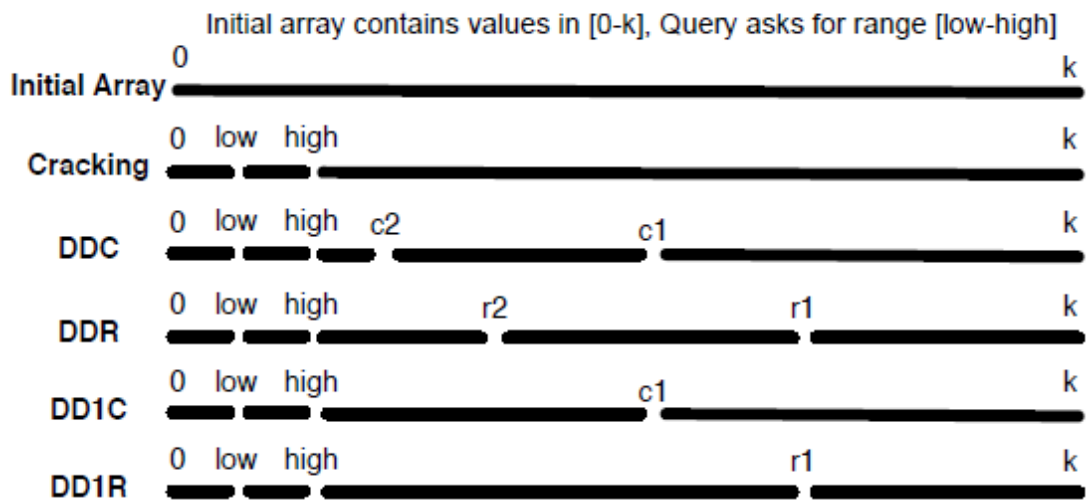

**Figure 4 shows cracking algorithms in action.[1]**

## Algorithm MDD1R

This is extension of DD1R using materialization concept. The algorithm works like DD1R, with the difference being that it does not perform the final cracking step based on the query bounds. Instead, it materializes the result in a new array.

In general, there are two bounds that define a range request in a select operator fall in two different pieces of an already cracked column. MDD1R handles these two pieces independently; it first operates solely on the leftmost piece intersecting with the query range, and then on the rightmost piece, introducing one random crack per piece. In addition, notice that the extra materialization is only partial, i.e., the middle qualifying pieces which are not cracked are returned as a view, while only any qualifying tuples from the end pieces need to be materialized. This highlights the fact that MDD1R does not forgo its query-driven character, even while it Eschews query-based cracking per se; it still uses the query bounds to decide where to perform its random cracking actions. In other words, the choice of the pivots is random, but the choice of the pieces of the array to be cracked is query-driven. It lacks only one weak point and that is still need to reduce the initialization cost and optimization as original cracking. [3]

## Algorithm PMDD1R

P stands for progressive cracking; in this the idea of incremental indexing is extended even at the individual cracking steps themselves. PMDD1R completes each cracking operation incrementally, in several partial steps; a physical reorganization action is completed by a sequence of queries, instead of just a single one. In design of progressive cracking, a restriction on the number of physical reorganization actions a single query can perform on a given piece of an array; in particular, control is introduced on number of swaps performed to change the position of the tuples. The resulting algorithm is even more lightweight than MDD1R; like MDD1R, it also tries to introduce a single random crack per piece (at most two cracks per query) and materializes part of the result when necessary. The difference of PMDD1R is that it only gradually completes the random crack, as more and more queries touch (want to crack) the same piece of the column.

# Experimental Analysis

The experiments for the above were performed in C++, using the C++ Standard Template Library for the cracker indices. All experiments ran on an 8-core hyper-threaded machine (2 Intel E5620 @2.4GHz) with 24GB RAM running Centos OS 5.5 (64-bit). As in past adaptive indexing work, experiments are all main-memory resident, targeting modern main-memory column-store systems. It used several synthetic workloads along with real workload from the scientific domain.
Results of experiment are based on standalone running above defined algorithms and by using flip-flop method i.e. switching from stochastic cracking to original cracking for all pieces in a column which become smaller than L1 cache; within the cache the cracking costs are minimized.

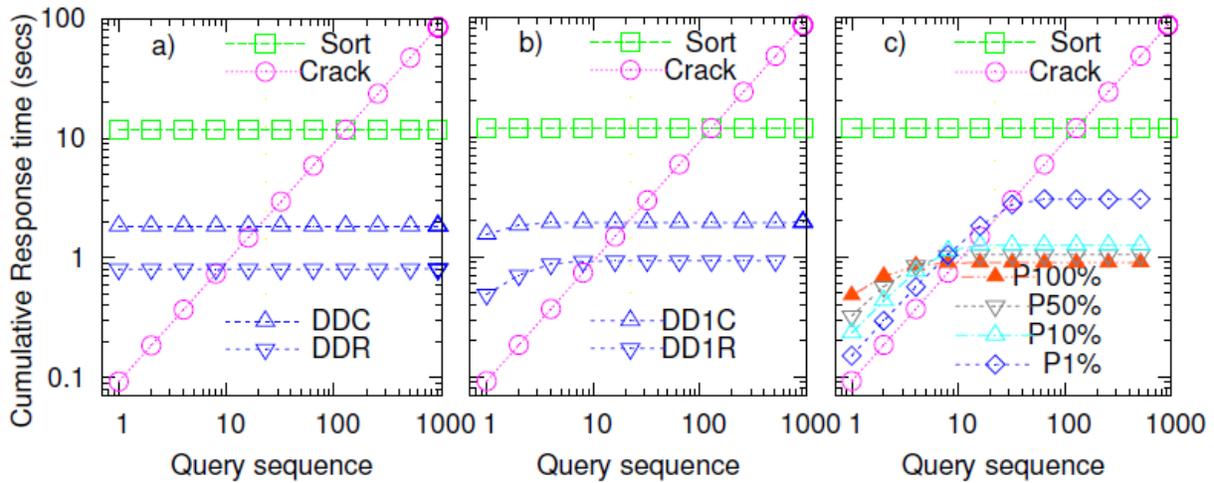

**Figure 5 shows Improving sequential workload via Stochastic Cracking.[1]**

Though it showed that stochastic cracking easily adapt any workload either random or non random but it failed to deliver performance in random workload as original cracking gives while remarkably improving its performance in sequential workload which original cracking is lacking.

|  | Random Workload selectivity % | | | | | Sequential Workload selectivity% | | | | |
|---|---|---|---|---|---|---|---|---|---|---|
| Algor. | $10^{-7}$ | $10^{-2}$ | 10 | 50 | Rand | $10^{-7}$ | $10^{-2}$ | 10 | 50 | Rand |
| Scan | 360 | 360 | 500 | 628 | 550 | 125 | 125 | 260 | 550 | 410 |
| Sort | 11.8 | 11.8 | 11.8 | 11.8 | 11.8 | 11.8 | 11.8 | 11.8 | 11.8 | 11.8 |
| Crack | 6.1 | 6.0 | 5.7 | 5.9 | 5.9 | 92 | 96 | 108 | 103 | 6 |
| DD1R | 6.5 | 6.5 | 6.4 | 6.4 | 6.4 | 0.9 | 0.9 | 1.1 | 1.5 | 5.9 |
| P10% | 8.6 | 8.6 | 10.3 | 10.3 | 10.3 | 1 | 1 | 1.9 | 3.4 | 9.1 |

**Figure 6 shows comparison for random and sequential workload.[1].**

## Solution: Optimization for Stochastic cracking

The notion of optimization stochastic cracking towards Robust Adaptive Indexing in Main-Memory Column-Stores can be achieved in two ways either by combining strength of algorithms defined above for stochastic database cracking via dynamic components that decides which algorithm to choose on the fly or by reducing of the initialization cost to make stochastic cracking even more transparent to the user, especially for queries that initiate a workload change and hence incur a higher cost.

## Proposed Algorithm

Copy the column required say ACRK in two pair array i.e. bit value set as 1 indicating corresponding value and values itself .Suppose our first select range query comes, this algorithm will show how to sort and crack it.

THREEPIECECRACKALGORITHM (low, high)
1. //here low and high are values given in range queries, n is length of ACRK
2. Create an arr1 =ACRK[size]//this time single empty array
3. Set chckpos =point at first position of ACRK
4. Set fpos = first position of aar1;
5. Set lpos = last position of arr2;
6. For (i=0;i<=n-1;i++)
7. { if (val.chkpos <low )
8. {value.fpos=val.chkpos ;
9. Set bitval=0;
10. Chkpos++;}
11. if (val.chkpos>high )
12. {value.lpos=val.chkpos;
13. Set bitval=0;
14. Chkpos ++;}
15. if (val.chkpos>=low and val.chkpos<= high)
16. { chkpos ++;
17. }
18. }
19. //this time bit val= =1 is considered and respective values are copied to new   //array
20. fpos++;
21. while (bit.val= =1 )
22. { Set value.fpos=value.chkpos
23. fpos++;
24. }

From the above algorithm we get sorted and cracked pieces, now delete the column ACRK. Next step is Choosing the piece having maximum no. of elements and apply original cracking on the basis of randomly chosen  no. from that piece resulting into 2 more pieces  for future benefits defined in [2][9-10].

# Summary and Future work

The given algorithm is proposed for optimization purpose of stochastic cracking so that it could attain a balance between sequential and random workload. Though the first steps seemed to be tedious but it for sure reduce the initialization cost as it completely reduced compare and swap process. Secondly it combined the two piece cracking algorithm for further cracking on basis of random number chosen number for the piece having maximum number of elements among other pieces obtained from first time cracking. The notion is to reduce the initialization cost for the purpose introduced in [1].  The above given algorithm can be implemented by making a benchmark using either c or c++ .Though the database we are considering is somewhat more than terabyte or so , for ease of implementation SKY server database could be used.
For motive of comparing performance between both cracking and newly designed cracking TPC-H benchmark could be used as to have an idea of this change done so far.
This is newly introduced part of big picture defined in [2] [10], everyone is welcome to further enhance the notion defined above for subsequent benefits designed prudently.

# REFERENCES

[1]  By  Felix Halim, Stratos Idreos, Panagiotis Karras, Roland H. C. Yap  "Stochastic Database Cracking: Towards Robust Adaptive Indexing in Main-Memory Column-Stores" ,2012

[2]   S. Idreos, S. Manegold, H. Kuno, and G. Graefe." Merging what's cracked,  cracking what's merged:Adaptive indexing in main-memory column-stores". PVLDB, 4(9):585–597, 2011.

[3]  By Stratos Idreos" Database Cracking:Towards Auto-tuning Database Kernels", 2010

[4]  G. Graefe and H. Kuno." Adaptive indexing for relational keys." SMDB, pages 69–74, 2010

[5]  P. Boncz, A. Wilschut, and M. Kersten. Flattening an Object Algebra to Provide Performance. In Proc. Of the IEEE Int'l. Conf. on Data Engineering, 1998.

[6]  S. Idreos, M. L. Kersten, and S. Manegold. "Database cracking. CIDR", pages 68–78, 2007

[7]  Peter Boncz (CWI)  Adapted from VLDB "Column-Oriented Database Systems "2009 Tutorial Column-Oriented Database Systems with Daniel Abadi (Yale) Stavros Harizopuolos (HP Labs)

[8]  By Martin Kersten, Stefan Manegold, joerd Mullender "The Database Architectures Research Group at CWI "

[9]  S. Idreos, M. L. Kersten, and S. Manegold." Self-organizing tuple reconstruction in column Stores". In  SIGMOD, pages 297–308, 2009

[10] By Stratos Idreos, Martin Kersten and Stefan Manegold CWI Amsterdam, The Netherland "database cracking ppts " 2010

## READING REFERENCES

[A] Goetz Graefe: Sorting and indexing with partitioned B-trees. CIDR 2003.

[B] M. Stonebraker, D. Abadi, A. Batkin, X. Chen, M. Cherniack, M. Ferreira, E. Lau, A. Lin, S. Madden, E.O'Neil, P. O'Neil, A. Rasin, N. Tran, and S. Zdonik. C-store: A column oriented dbms. In Proc. of the Int'l. Conf. on Very Large Data Bases, 2005.

[C] Stratos Idreos, Stefan Manegold and Goetz Graefe Ppts on "adaptive indexing in modern databases "

[D] Goetz Graefe:" Sorting and indexing with partitioned B-trees". CIDR 2003

[E] Goetz Graefe: Implementing sorting in database systems. ACM Comput. Surv. 38(3): (2006).

[F] Milena Ivanova, Martin L. Kersten, Niels Nes: Self-organizing strategies for a column-store database.EDBT 2008: 157-168

[G] Oguzhan Ozmen,Mustafa Uysal,M. Hossein Sheikh Attar "Storage Workload Estimation for Database ManagementSystems